\documentclass[apj]{emulateapj}

\input{epsf}

\def\***#1{\textsf{#1}}

\def\M324{\ensuremath{M_{g,324}}}

\def\p0{\phantom{0}}

\begin{document}

\shorttitle{A GALAXY CLUSTER BEHIND M31}
\shortauthors{KOTOV, TRUDOLYUBOV, \& VESTRAND}
\slugcomment{The Astrophysical Journal, accepted}

\title{A CLUSTER OF GALAXIES HIDING BEHIND M31: \emph{XMM-Newton} OBSERVATIONS OF 
RXJ0046.4+4204}

\author{Oleg Kotov\altaffilmark{1,2,3}, Sergey Trudolyubov\altaffilmark{1,2,4}, 
and W. Thomas Vestrand\altaffilmark{1}}

\altaffiltext{1}{NIS-2, Space and Remote Sciences Group, Los Alamos National Laboratory, Los Alamos, NM 87545}
\altaffiltext{2}{Space Research Institute, Russian Academy of Sciences, Moscow, Russia}
\altaffiltext{3}{Harvard-Smithsonian Center for Astrophysics, 60 Garden St., Cambridge, MA 02138}
\altaffiltext{4}{Institute of Geophysics and Planetary Physics, University of California, Riverside, CA 92521}

\begin{abstract}
We report on our serendipitous discovery with the XMM-Newton Observatory of
a luminous X-ray emitting cluster of galaxies that is located behind the
Andromeda galaxy (M31). X-ray emission from the cluster was detected
previously by ROSAT, and cataloged as RX J0046.4+4204, but it was not
recognized as a galaxy cluster.  The much greater sensitivity of our
XMM-Newton observations revealed diffuse x-ray emission that extends at
least $5\arcmin$ and has a surface brightness profile that is well fit   
by the $\alpha$-$\beta$ model  with $\beta = 0.70 \pm 0.08$, a
core radius $r_c = 56 \arcsec \pm 16 \arcsec$, and $\alpha = 1.54 \pm 0.25$.
A joint global spectral fit of the EPIC/MOS1, MOS2, and PN observations 
with  Mewe-Kaastra-Liedahl plasma emission model gives a cluster temperature of 
$5.5 \pm 0.5$ keV. The observed spectra also
show high significance iron emission lines that yield a measured cluster
redshift of z = 0.290 with a 2$\%$ accuracy.  For a cosmological model with
$H_0$ = 71 km s$^{-1}$ Mpc$^{-1}$, $\Omega_M$ = 0.3 and $\Omega_{\Lambda}$ =
0.7 we derive a bolometric luminosity of  $L_x=(8.4 \pm 0.5)\times 10^{44}$ erg/s.
This discovery of a cluster behind M31 demonstrates the utility of x-ray surveys for 
finding rich clusters of galaxies, even in directions of heavy optical extinction.
\end{abstract}



\section{INTRODUCTION}
 
Galaxy clusters are the largest gravitationally bound structures in the universe.
The evolution of cluster number density of a given mass  is sensitive to specific
cosmological scenarios (e.g. \cite{1974ApJ...187..425P}). So observations of galaxy clusters 
are an important tool for constraining fundamental cosmological parameters. 

Due to the fact that $15\%$ of the total cluster mass (e.g. \cite{1997MNRAS.292..289E})
is in the form of hot diffuse plasma emitting at X-ray band via thermal bremsstrahlung  
\citep{Sarazin88}, galaxy clusters are among the most luminous objects in X-ray band.
It makes X-ray selection an efficient means for constructing samples of galaxy clusters (see review 
by \cite{2002ARA&A..40..539R}) .
X-ray selection has the advantage that the measurable X-ray luminosity and temperature are 
correlated with the cluster mass.
  
Further, X-ray selection is useful for studying regions where optical searches 
are complicated because of dust extinction and heavy stellar 
confusion. X-rays are much less affected by extinction than optical photons 
and X-ray selection is almost free from  source confusion problems \citep{2002ApJ...580..774E}.
Conducting X-ray selection based on {\em ROSAT} data at low Galactic latitude, 
\cite{2002ApJ...580..774E} were able to detect 137 galaxy clusters ,
$70\%$ of which were new discoveries.

With a new era of {\em XMM-Newton} and {\em Chandra} observatories with their large effective areas
and wide energy ranges $0.3-10$keV, the capability of X-ray selection increased.
During an {\em XMM-Newton} observation of the galactic supernova remnant G21.5-09 located close to
the Galactic Plane, \cite{Nevalainen01} detected a new galaxy cluster. Using only {XMM-Newton} data,
they measured  cluster redshift $z = 0.1 $ to $1\%$ precision that is especially important in
regions with such strong optical source confusion, where the optical
redshift measurements of galaxies are difficult .

Here we present new XMM-Newton observations of the source RX J0046.4+4202
that indicate it is a high redshift cluster located behind M31.  
RX J0046.4+4204 was detected during the first and the second deep ROSAT RSPC surveys of 
M31 performed in June 1991 and July/August 1992 respectively \citep{Supper01}.
Based on a comparison of the first and the second surveys, RX J0046.4+4204 was classified as
a potentially long term variable source.
Our analysis of the data obtained with XMM-Newton revealed spatially 
extended emission, up to at least $5\arcmin$, from  RX J0046.4+4204. 
The observed spectra show iron emission lines that yield a measured
redshift of  z = 0.290 with a 2$\%$ accuracy. All these facts
combined with optical image from Digitized Sky Survey allow us to
conclude that  RX J0046.4+4204 is actually a distant galaxy cluster.

In this paper, we assume the $\Omega_M = 0.3$ and $\Omega_{\Lambda} = 0.7$
cosmology with the Hubble constant of $H_0$ = 71 km/s/Mpc.
For the defined above cosmology and the measured
redshift of  z = 0.290, the angular size of $1\arcmin$ corresponds the
physical size of 257 kpc.
Statistical uncertainties are quoted at the 90$\%$ confidence level unless there is a 
statement saying otherwise.

\section{OBSERVATIONS AND DATA REDUCTION}
\begin{figure*}[htb]
\centerline{
\includegraphics[width=0.48\linewidth]{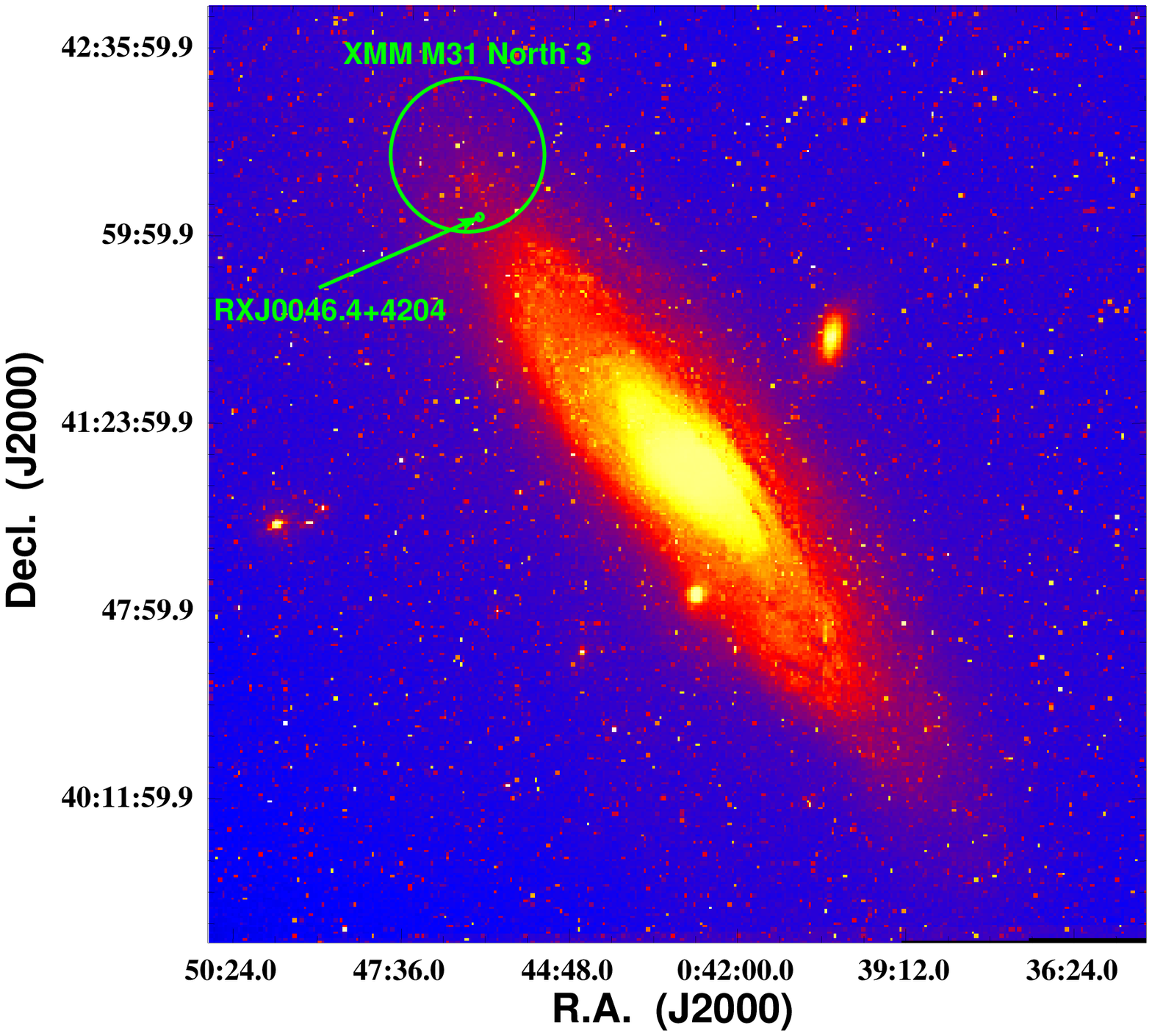}%
\hfill%
\includegraphics[width=0.51\linewidth]{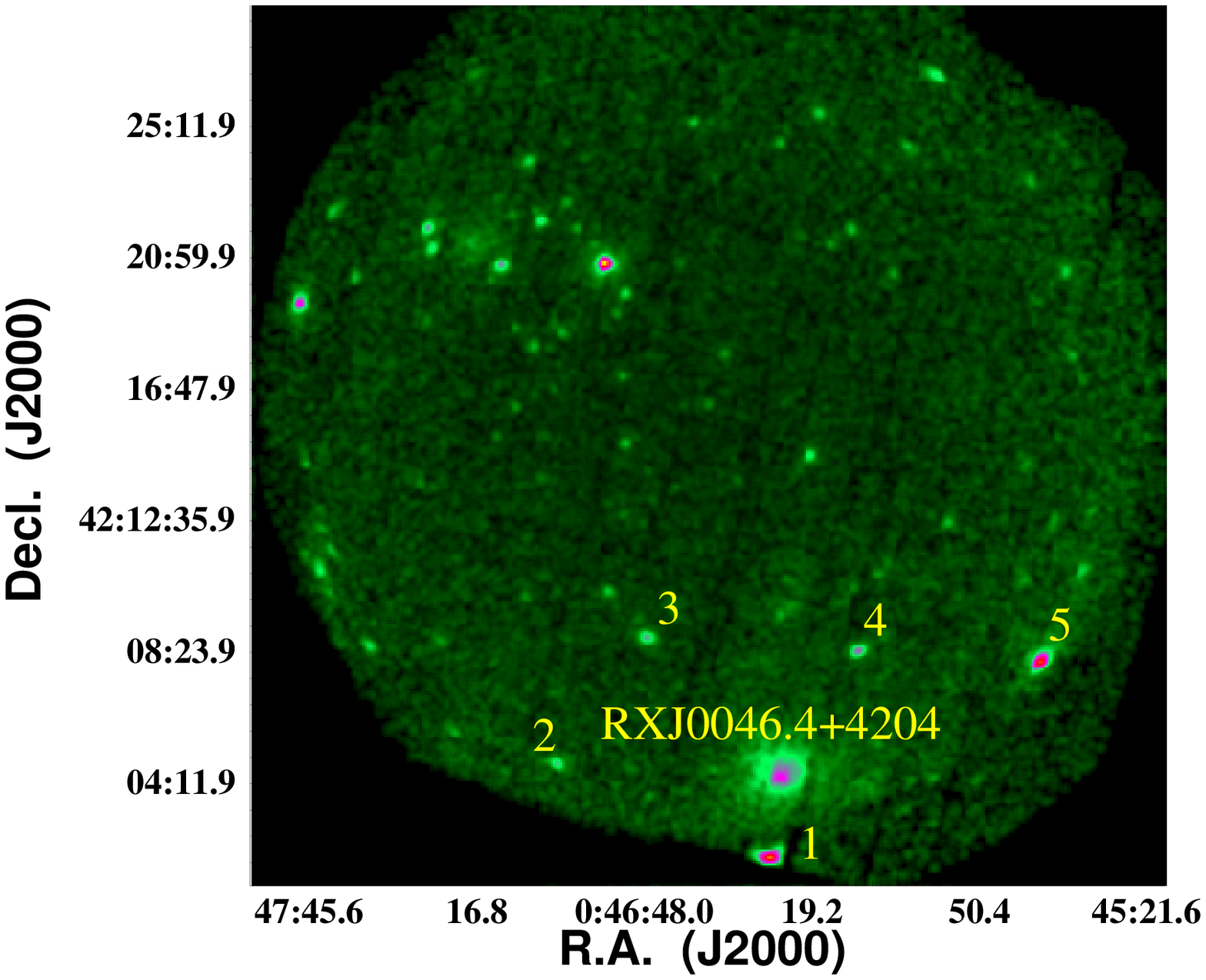}%
}
\caption{
{\em Left:}Optical image of M31 from Digitized Sky Survey 
with {\emph XMM-Newton} FOV show as a circle for M31 North 3 field.
{\em Right:} The combined MOS1-MOS2-PN vignetting-corrected image of 
the XMM North 3 Field of M31 in $0.8 - 2.5$ keV energy band, 
square root intensity scale. 
\label{image}}
\end{figure*}

\begin{figure}
\epsfxsize=8.5cm
\epsffile{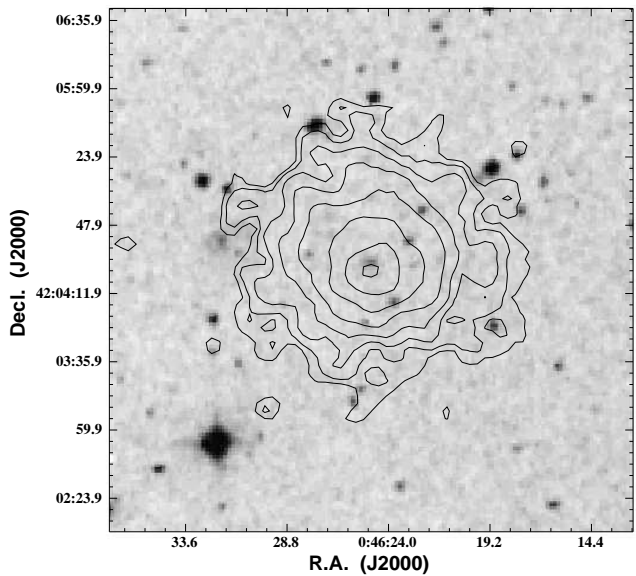}
\caption[]{The Palomar Digitized Sky Survey optical image with overlaid brightness contours  made from 0.8-2.5keV band MOS2/PN image. The X-ray image was smoothed with a $4\arcsec$ Gaussian kernel.
The contour levels are based on the background noise going up in
logarithmic steps. MOS1 image was not used for construction of the contours because the
center of the cluster falls at the edge of a CCD in MOS1 camera. 
The angular size of the largest contour corresponds
to $\sim0.7$Mpc at $z=0.29$.
\label{optics}} 
\end{figure}
In the following analysis, we use the data from {\em XMM-Newton} 
observation of the XMM North 3 Field of M31 centered at 
${\rm RA} = 00^{h} 46^{m} 38^{s}.00$; ${\rm Decl} = +42\arcdeg 16\arcmin 20.0\arcsec$.
Fig.~\ref{image} ({\em Left})  shows optical image of M31 from Digitized Sky Survey  
with {\emph XMM-Newton} FOV shown as a circle for M31 North 3 field.
The {\em XMM-Newton} observation was performed on 2002 June 29 as a 
part of the Guaranteed Time Program (PI: K.O. Mason). An 
analysis of the complete North 3 field  is presented in \cite{astro-ph/0401227}.

In current analysis, we used the data from two
EPIC-MOS  detectors \citep{2001A&A...365L..27T} 
and the EPIC-PN detector \citep{2001A&A...365L..18S}.   
The EPIC data was reduced using the standard {\em XMM-Newton} Science Analysis 
System (SAS v6.1.0)\footnote{See http://xmm.vilspa.esa.es/external/xmm\_sw\_cal/sas.shtml}.
We used the calibration database with all updates available prior to January, 2005.
Only X-ray events corresponding to patterns 0-12  for MOS detectors and
patterns 0-4 for PN detector were selected. All known bad pixels were excluded.

The EPIC background is highly variable and only its quiescent component can be
accurately modeled. To detect and exclude the periods of high flaring background, 
we produced the light curves  for each EPIC detector  showing count rate in the 2.0-15.0keV energy 
band from the whole field of view but with  detected  sources masked.
The light curves  were binned to 200~s time resolution. We screened the EPIC data 
to recursively  exclude time  intervals with  the deviation of the count rate 
exceeding a $2\sigma$ threshold from the average.
Experiments with different choices of  energy bands and flare detection thresholds have 
shown that our choice was close to optimal. The remaining good exposure time was $\sim 24$ ks 
for the EPIC-PN, $\sim 41$ ks for the EPIC-MOS1, and $\sim 43$ ks for the EPIC-MOS2.

To account for strong \emph{XMM} mirror vignetting, we used an approach
proposed by \cite{Arnaud01}. Each photon was assigned a weight
proportional to inverse vignetting and these weights were then used in
computing images and spectra. This was done using the SAS tool
\emph{evigweight}.

Background modeling in our analysis was implemented following the
double-subtraction method of \cite{2002A&A...390...27A}.  The first step
of this method  is to subtract the particle-induced background component. 
This component can be estimated from a set of XMM
observations with the filter wheel closed (so called ``closed data'').
We compiled the closed dataset from public observations available in the
XMM data archive; these data were reduced following identical steps as
the science observations. The closed background was adjusted to the
cluster observations using the observed flux in the 10--15~keV band
outside the field of view. The scaling factors are 1.04, 1.05, and 1.01 for 
MOS1, MOS2, and PN respectively.
The second step is to determine the cosmic X-ray
background (CXB) component. Its spatial distribution should be flat
because vignetting correction is already applied. Therefore, the CXB
component can be measured in the source-free regions of the field of
view at large radii from the cluster center (see \S~3.1 and 3.2 below).

Finally, we  applied  a correction for photons registered during  the CCD readouts, 
so called out-of-time events, to the EPIC-PN data.

\section{Results}

The combined MOS1, MOS2, and PN image of the XMM North 3 Field of M31
in the $0.8 - 2.5$ keV energy band,  corrected for the effects of instrumental 
vignetting, is shown in Figure~\ref{image} ({\em Right}). The raw image was 
convolved with a Gaussian function with spatial scale $\sigma= 4 ''$.
We define the cluster center to be at the location of the X-ray surface brightness 
peak of the cluster emission in the $0.8 - 2.5$ keV energy band, $\alpha =
00^{h}46^{m}24^{s}.8$   $\delta = +42\arcdeg04\arcmin26\arcsec$ (J
2000),  with an estimated uncertainty radius of $6 ''$ ($90\%$ CL),
determined by the wavelet decomposition algorithm of \cite{1998ApJ...502..558V}.
The Palomar Digitized Sky Survey optical 
image  shows no extended optical counterpart for RX J0046.4+4204 (see Fig.~\ref{optics}).

During the following spectral and spatial analysis, we excluded all detectable point
sources from the data. The sources were detected separately in the ``optimal'' 0.3-3~keV,
``soft'' 0.3-0.8~keV, and ``hard'' 2.0-6.0~keV energy bands using the
task {\em wvdecomp} of the ZHTOOLS package
\footnote{Seehttp://hea-www.harvard.edu/$\thicksim$alexey/zhtools/}.
Detected point sources were masked with circles of 80\% PSF power radii.

\subsection{Spatial Analysis}

\begin{deluxetable}{lcc}
\tablecolumns{3} 
\tablewidth{0pc} 
\tablecaption{Results of Spatial Analysis. Combined EPIC-PN, MOS1 and MOS2 data, 
$0.8 - 2.5$ keV energy  range. Parameter errors quoted are $90\%$ confidence limits 
\label{spatial_analysis}.}
\tablehead{ 
\colhead{Parameters}      &    \colhead{$\beta$ fit}  & \colhead{$\alpha-\beta$ fit}} 
\startdata 
      $\alpha$            &         \nodata           &    $1.54 \pm 0.25$  \\
      $\beta$             &     $0.60  \pm 0.03$      &    $0.70 \pm 0.08$  \\
   $r_{c}(\arcsec)$       &     $20.2 \pm 2.8$        &    $56  \pm   16 $  \\
$r_{c}$(kpc for $z$=0.290)&     $87 \pm 12$           &    $240  \pm  69 $  \\ 

      $\chi^2$/dof    &      156.7(119)           &       139.4(118)           
\enddata
\end{deluxetable}

Spatial  analysis of the cluster emission was performed in the $0.8 - 2.5$ keV 
energy band using all detectors. Experiments with different energy bands showed
that the signal to noise ratio was close to optimal for the chosen
band. We used a pixel size of $4 \arcsec$  in our spatial analysis.
The image for each camera was corrected for vignetting. The PN image was
corrected for out-of-time events.We subtracted the particle background component 
from the images as described in \S~2.

We extracted the azimuthally averaged surface brightness profiles
centered on the X-ray surface brightness peak, excluding the CCD gaps and 
circles around the point sources.  The profiles were logarithmically
binned with a step of $\Delta r = 0.1 r $.  A logarithmic radial binning
approximately preserves the signal to noise ratio in annuli until the
background becomes comparable to the signal.  For the extracted profiles,
the chosen step keeps the signal to noise ratio above 3 in annuli until $\sim 5\arcmin$.
The obtained profiles were used to derive
the parameters of the spatial distribution of the ICM, 
cluster fluxes, and the level of the CXB component.

The cluster surface brightness profiles are often modeled with the so
called $\beta$-model, $n_e^2 \propto (1+r^2/r_c^2)^{-3\beta}$ or $S_x
\propto (1+r^2/r_c^2)^{-3\beta+0.5}$ \citep{Cavaliere76}. 
However, this model poorly describes
clusters with sharply peaked surface bigness profiles  related to
the radiative cooling of the ICM in the cluster centers.

We used the simple  modification of the $\beta$-model 
\begin{equation}\label{eq:alpha-beta}
n_e^2 \propto \frac{(r/r_c)^{-\alpha}}{(1+r^2/r_c^2)^{3\beta-\alpha/2}},
 \end{equation} 
which allows to model a power law-type emission excess in the cluster centers. The similar models were also used by \cite{Pratt02} and \cite{astro-ph/0507092}.

 The model for the observed surface brightness profiles can be obtained
 by numerical integration of eq.(\ref{eq:alpha-beta}) along the line of
 sight, and convolution of the result with the XMM PSF\footnote{We used
   the latest available values of the King function parametrization of
   the MOS1,
   MOS2, PN PSF, see\\
   http://xmm.vilspa.esa.es/docs/documents/CAL-TN-0018-2-4.pdf}. 
 To represent the uniform sky X-ray background, we added a constant
 component to the model and treated it as a free parameter.  The values
 of $\alpha$, $\beta$, and $r_c$ were derived from the joint fit to the
 observed profiles in the MOS1,2 and PN cameras, with the overall
 normalizations and background levels fit independently for each camera.
 The obtained parameters for the $\alpha$-$\beta$ model 
 are summarized in Table~\ref{spatial_analysis}.

We obtained a $\beta$ value of $0.70 \pm 0.08$. 
 For a sample of local clusters, \cite{1999ApJ...525...47V} showed that
 if cooling flow regions were excluded from $\beta$-model fit, values of the 
 $\beta$ parameter  were distributed over a narrow range $0.7 \pm 0.1$. 
 So the value of the $\beta$ parameter we derived describes the typical distribution of the ICM.

The obtained value of the core radius ,  $r_c = (56 \pm 16)\arcsec$ (($240\pm 69$) kpc for $z$=0.290 ), is similar to the values of the $r_c$ parameter \cite{1999ApJ...525...47V} measured for a sample of local clusters.

For the value of $\alpha$ parameter we obtained  $\alpha=1.54 \pm 0.25$. 
 This value is consistent with values of the $\alpha$ parameter \cite{astro-ph/0507092} 
 derived for typical cooling-flow clusters using a sample of local clusters. 
 \cite{astro-ph/0507092}  in their analysis 
  used {\em  Chandra} data  where the PSF is not an issue, so their measurements of $\alpha$ 
  parameter are more direct and do not  depend on  the quality of a PSF model 
   as in our case. 

While the EPIC-pn PSF is azimuthally symmetric  and is well calibrated up to large offset angles in the 0.8-2.5 energy band (see Fig 1-6 in \cite{Kirsch05}),
the EPIC-MOS PSF is less reliable and seems not azimuthally symmetric. 
The asymmetry is not currently modeled in the SAS calibration.
To check how imperfect calibration of the EPIC-MOS PSF can
influence the obtained result, we fitted  the surface brightness profile of 
the cluster extracted from the EPIC-PN detector alone. 
The obtained best-fit values were  within our statistical uncertainties.

  We can quantify the strength of cooling flow calculating the fraction , 
  $f_{70\mbox{\scriptsize kpc}}$ of the count rate coming from a central region of 70kpc to  
  the total count rate \citep{1998MNRAS.298..416P}.  This  quantity exhibits 
  large dispersion, but in general, clusters
  with massive cooling flows  show  $f_{70\mbox{\scriptsize kpc}} > 15 \%$  \citep{1998MNRAS.298..416P}. 
  For a typical non-cooling-flows cluster $f_{70\mbox{\scriptsize kpc}}$ $\sim 6 \%$ 
  \citep{Markevitch98}.
  We estimated the count rate coming from the 70kpc central region using
   the best fit $\alpha$-$\beta$ model. The total count rate was calculated integrating the observed
   surface brightness profile within $1.4$Mpc radius.
   We obtained $f_{70\mbox{\scriptsize kpc}}=18\%$.

\begin{figure*}[htb]
\centerline{
\includegraphics[width=0.5\linewidth]{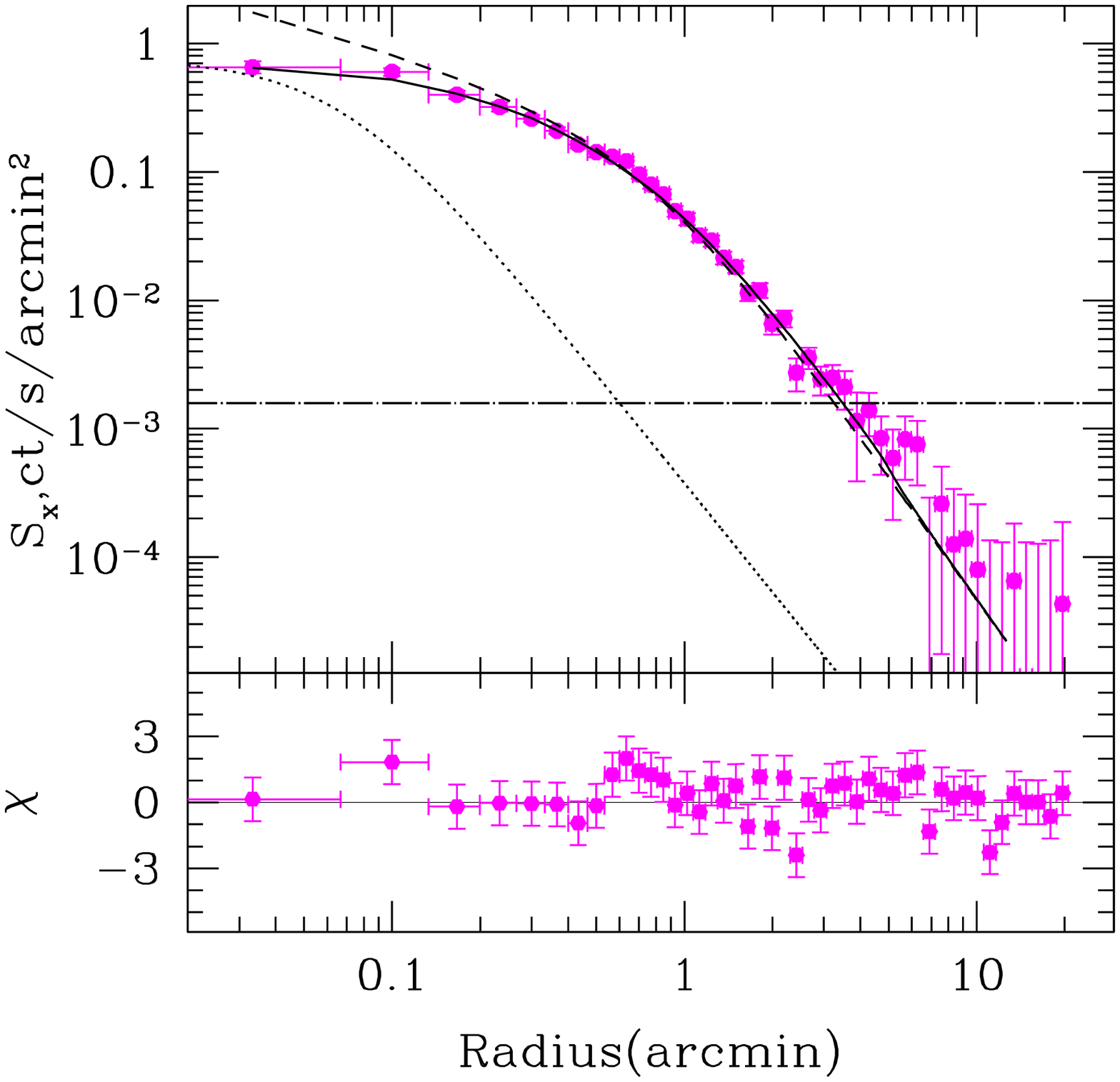}
\hfill
\includegraphics[width=0.5\linewidth]{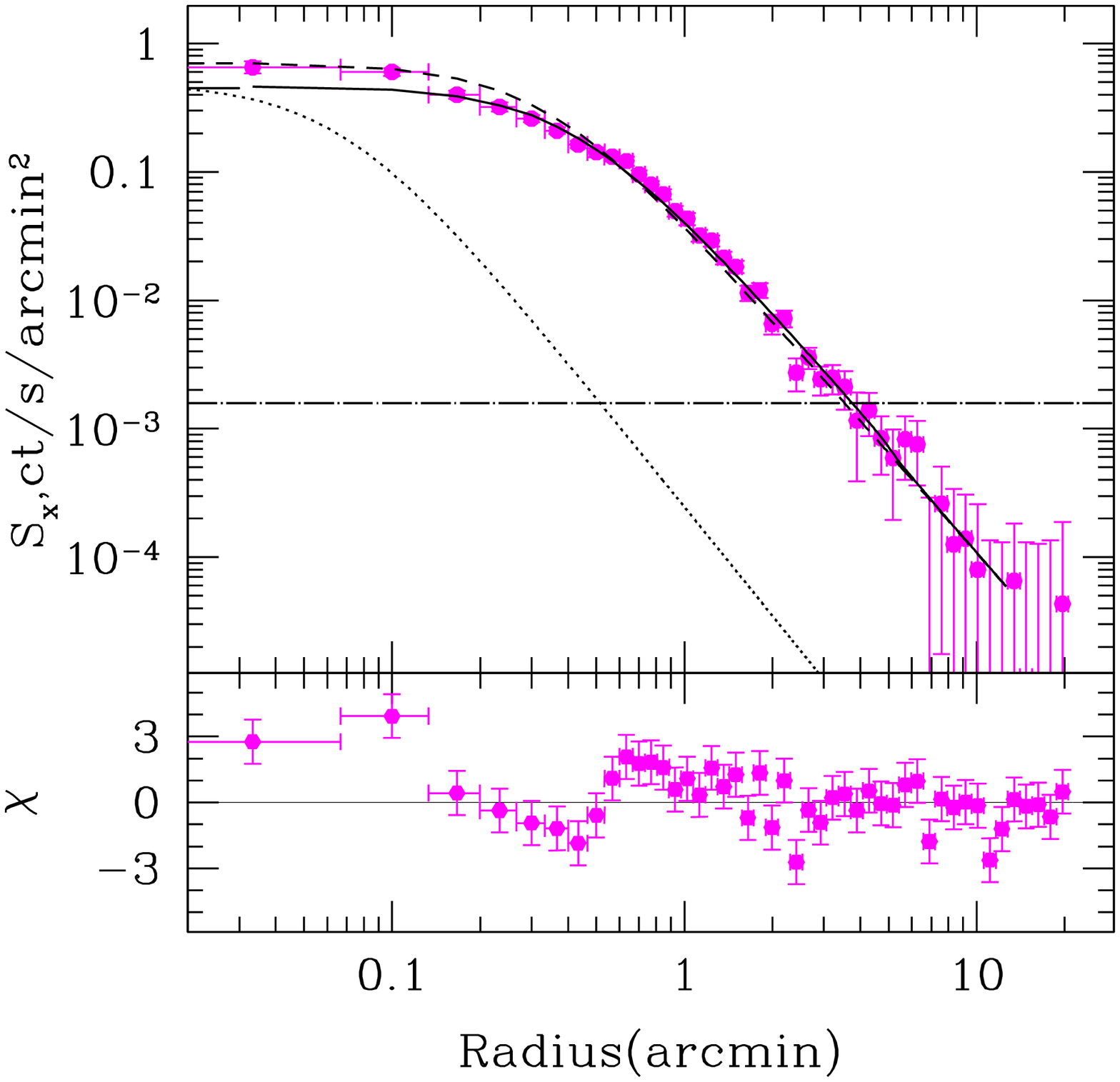}
}
\caption{{\em Left:} Upper panel: The combined MOS1-MOS2-PN surface  brightness profile of RX J0046.4+4204 in  the $0.8 - 2.5$ keV energy band. Black solid line shows $\alpha$-$\beta$ model best fit convolved with the XMM PSF. Black dashed line shows $\alpha$-$\beta$ model best fit. Black dotted line shows the XMM PSF. Black dotted-dashed line shows the level of the cosmic X-ray background component. Lower panel: The residual between the data and the best-fit model in terms of sigmas. 
{\em Right:} The same with the standard $\beta$-model best fit. 
\label{spacial_fig}}
\end{figure*}

  The combined MOS1-MOS2-PN  surface brightness profile along with the best fit $\alpha$-$\beta$ model
  are shown in Fig.~\ref{spacial_fig}. From Fig.~\ref{spacial_fig} one can see how 
  the XMM PSF can flatten a peaked profile.  
  For comparison, we also fit the surface brightness profiles 
  by the standard $\beta$-model  setting $\alpha=0$.
  The obtained parameters are  summarized in Table~\ref{spatial_analysis}.
 
\subsection{Global Spectrum}

\begin{figure*}[htb]
\centerline{
\includegraphics[width=0.5\linewidth]{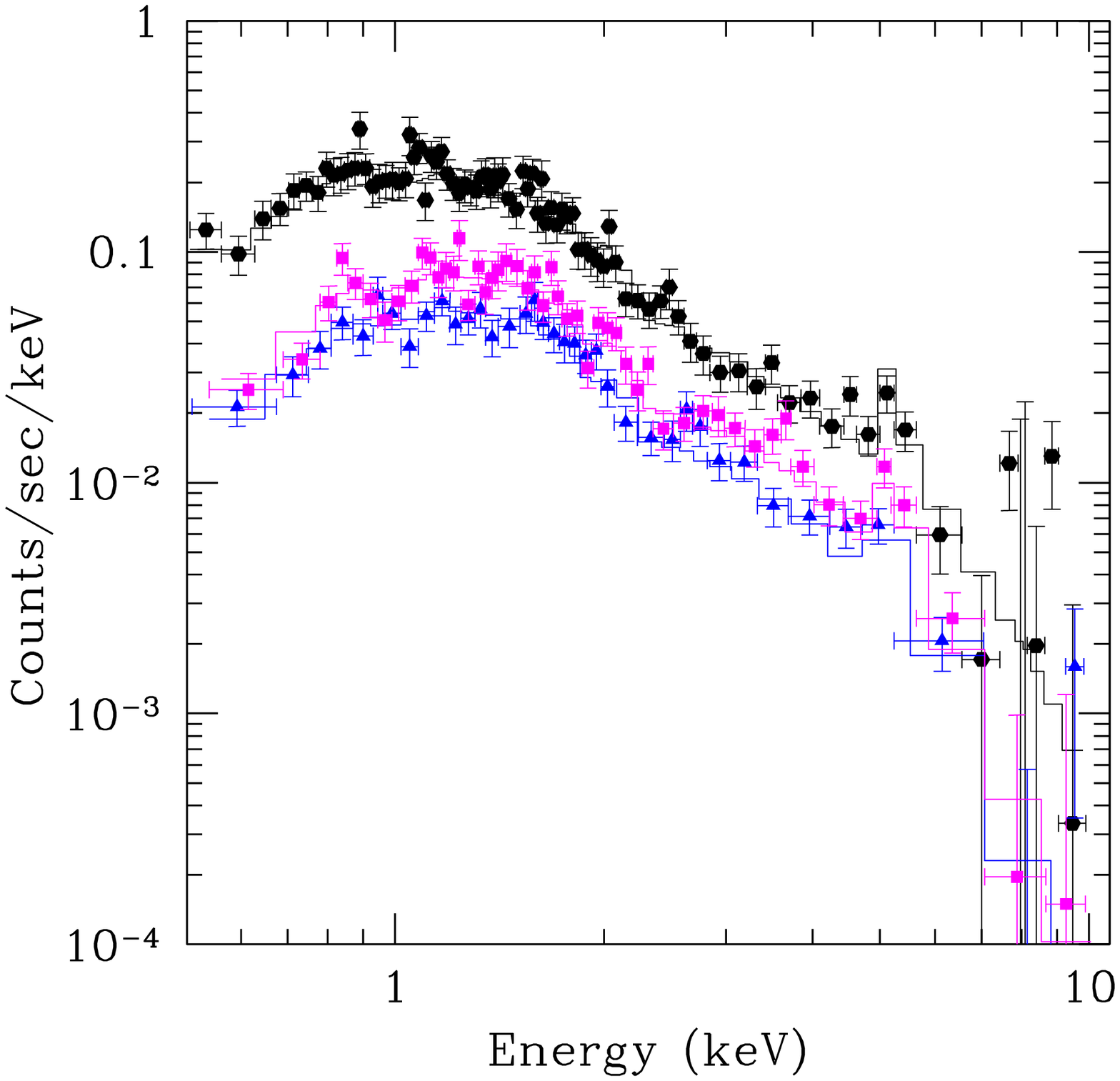}
\hfill
\includegraphics[width=0.5\linewidth]{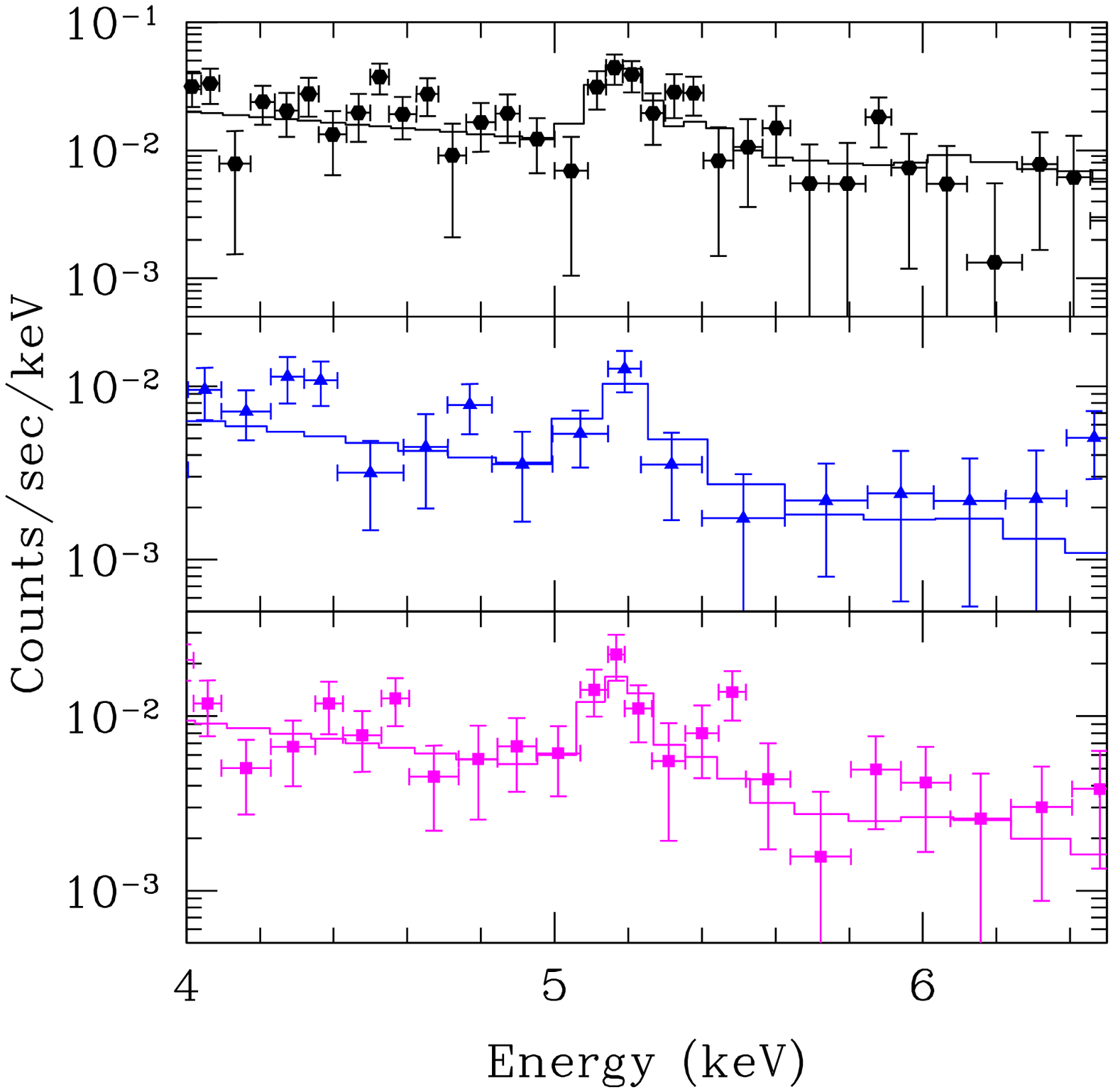}
}
\caption{{\em Left:} Count spectra of RX J0046.4+4204 
taken with the {\em XMM-Newton}/EPIC-PN ({\em black circles}),
MOS1 ({\em blue triangles}) and MOS2 detectors ({\em magenta squares}).
The corresponding  best-fit spectral models (absorbed red-shifted  
{\em Mekal} model) convolved with instrumental responses are shown as {\em black},
{\em red} and{\em blue} solid lines.
{\em Right:} Expanded view of the 4.0-7.0keV energy band. Upper, Middle and Lower panels show
PN, MOS1 and MOS2 data.  A red-shifted iron emission line feature is clearly evident.
\label{spectrum_fig}}
\end{figure*}
For our spectral analysis, we used the screened vignetting-corrected  data in the $0.5 - 10.0$ keV
energy band from all cameras. The spectra of the cluster were extracted from a circular region with 
angular radius of $3.1\arcmin$ (0.8Mpc for $z$=0.290)  for all EPIC data. All detected contaminating point-like sources were 
excluded from the source and background extraction regions.
We subtracted the particle background component from the images as described in \S~2.
To estimate the remaining CXB component, we extracted the spectra from a circular region
with angular radius  of $9.0\arcmin$  centered at the on-axis position  ,
but  a part of the region falling into the $9.65\arcmin$ (2.5Mpc for
$z$=0.290) circle centered at the cluster center 
was excluded.
The response matrices and effective area
files were generated by the standard SAS tasks.  Because the data were
previously vignetting corrected, the effective area files were created
for the on-axis position using the routine {\em arfgen}.  The response
matrices were generated in the spectrum extraction region via {\em rmfgen}.

\begin{deluxetable*}{ccccc}
\tablecolumns{3} 
\tablewidth{0pc} 
\tablecaption{Results of Spectral Analysis. EPIC-PN, MOS1 and MOS2 data, $0.5 - 10.0$ keV 
energy range. Spectral extraction radius is $3.1\arcmin$. Parameter errors quoted are $90\%$ confidence limits \label{spectral_analysis}.}
\tablehead{ \colhead{Parameters} & \colhead{MOS1}       &    \colhead{MOS2}  & \colhead{PN}  & \colhead{Combined}}
\startdata 
$N_{H}(10^{22} cm^{-2})$  &  $0.21\pm^{0.04}_{0.03}$    &  $0.26\pm^{0.03}_{0.03}$    &  $0.21\pm^{0.02}_{0.02}$       &  $0.22\pm^{0.02}_{0.02}$ \\
$kT(keV)$                 &  $6.4\pm^{1.4}_{1.1}$       &  $5.3\pm^{0.8}_{0.7}$        &  $5.1\pm^{0.7}_{0.5}$         &  $5.5\pm^{0.5}_{0.5}$    \\  
$Z_{\odot}$               &  $0.44\pm^{0.19}_{0.27}$    &  $0.72\pm^{0.20}_{0.19}$     &  $0.45\pm^{0.21}_{0.10}$      &  $0.57\pm^{0.15}_{0.13}$ \\
 $z$                      &  $0.296\pm^{0.011}_{0.023}$ &  $0.292\pm^{0.009}_{0.007}$  &  $0.287\pm^{0.008}_{0.008}$   & $0.290\pm^{0.005}_{0.005}$   
\enddata
\end{deluxetable*}

The source spectra were binned to have at least 30 counts in each
spectral bin and fit in XSPEC 11.3.0 \citep{Arnaud96} by 
the Mewe-Kaastra-Liedahl plasma emission model \citep{1985A&AS...62..197M}.
We used abundances from \cite{1989GeCoA..53..197A}. 
Galactic photoelectric absorption was accounted for using the WABS
model \citep{1983ApJ...270..119M}.
The spectra from the EPIC-PN (3677 counts), MOS1 (1666 counts) and 
MOS2 (2179 counts) detectors were fitted both jointly and
separately. For the joint fits, only spectral model normalizations were
allowed to vary independently.The results of both joint and separate spectral fitting of 
the EPIC-PN, MOS1 and MOS2 data are summarized in Table \ref{spectral_analysis}.

We obtained a redshift value of 0.290 with 2$\%$ 
accuracy. The redshift values estimated independently from EPIC-PN,  
MOS1 and MOS2 data are in good agreement within measurement errors.
For the value of hydrogen column density 
we obtained  $N_{H} = (2.2 \pm ^{0.2} _{0.2}) \times 10^{21}$ cm$^{-2}$ that is
significantly above the Galactic hydrogen column  in the direction of M31, $\sim 7   
\times 10 ^ {20} cm ^ {-2}$ \citep{1990ARA&A..28..215D}.
We checked that the values of absorption  obtained independently from EPIC-PN, MOS1 and MOS2
data are consistent within  the measurement errors.

The EPIC-PN, MOS1 and MOS2 
spectra, along with the best-fit spectral models, are shown in Figure~\ref{spectrum_fig}. 
Note that separate spectral fitting of the EPIC-PN, MOS1 and MOS2
data gives consistent values of the model parameters.

\subsection{Temperature profile }
\begin{figure}
\plotone{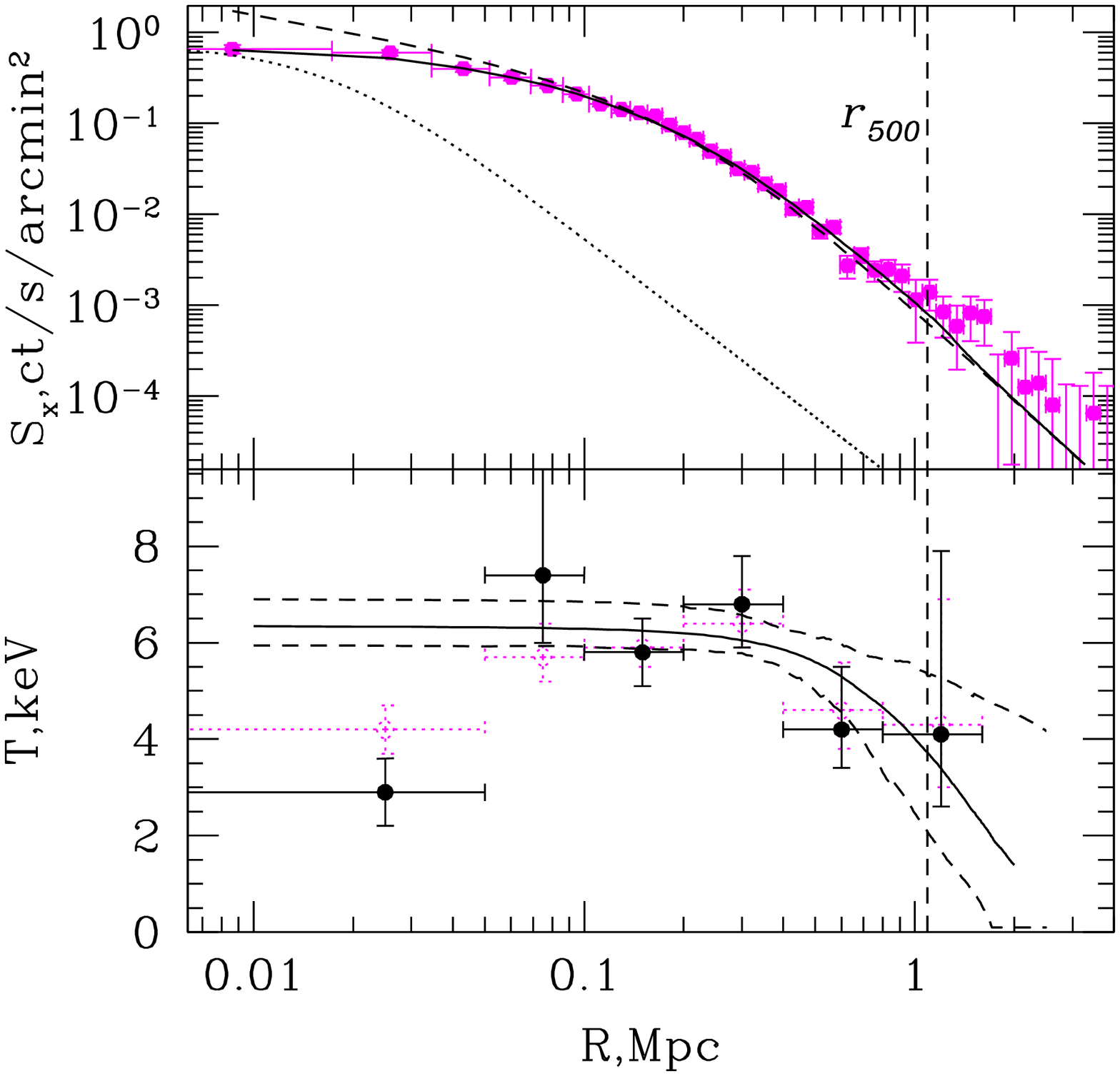}
\caption[]{Upper panel:  Same as Fig.~\ref{spacial_fig} ({Left upper panel}).
Lower panel:  The temperature profile of the cluster
 as a function of angular radius obtained from EPIC-PN/MOS1/MOS2 data.
 Solid black circles show the deconvolved projected temperature
profile.  For comparison, open circles shows the raw measurement from
the X-ray fit in the same annuli.The  error bars correspond to $68\%$
(1-$\sigma$) confidence limits.  Solid line shows the best-fit projected
  temperature profile and dashed lines correspond to its 68\% CL
  uncertainties.
\label{oneplot}} 
\end{figure}

To construct temperature profile, we extracted individual spectra 
in 5 annuli: $0-50$kpc , $50-100$kpc , $100-200$kpc, $200-400$kpc, and
$800-1600$kpc, centered on the position of the X-ray surface brightness peak using
data from all cameras.  After background subtraction, 
the number of counts (MOS1+MOS2+PN) in 0.5-10kev energy band in  1, 2, 3, 4, and 5 annuli was
 632, 1275, 2071, 2069, 1063, and 459 accordingly.

The correction for the XMM PSF effect was done using an approach of \cite{2004A&A...423...33P}.  
Using the best-fit $\alpha$-$\beta$ models of the cluster brightness and the XMM
PSF calibration, we calculated the redistribution matrix, $R_{ij}$,of
each temperature to each annulus which represents relative contribution
of emission from annulus $i$ to the observed flux in annulus $j$. 
The model spectrum, $S_j$, is then given by 
\begin{equation}
  \label{eq:t:mix}
  S_j=\sum R_{ij} S(T_i),
\end{equation}
where $T_i$ is the temperature in annulus $i$ and $S(T_i)$ is the
\emph{mekal} spectrum for this temperature. Fitting this model to the
observed spectra in all annuli simultaneously and treating all $T_i$ as
free parameters gives the deconvolved temperature profile. Unfortunately,
the statistic was poor to measure metallicity profile. 
So for all annuli we fixed the metallicity values at the best-fit
values obtained from the global spectrum. 
We checked that allowing the absorption and the metallicity to be freely fitted does not significantly
change the result. The values of redshift and absorption values were also fixed at the best-fit values.

We fitted  the observed temperature profiles by a 3-D temperature model :
\begin{equation}
  \label{eq:t:model}
T(r) = T_0/(1.+(x/0.6)^2)^{\gamma}, 	
\end{equation}
where $x=r/r_0$.
A similar model describes the temperature profiles for low redshift clusters 
\citep{astro-ph/0507092}. 
For local clusters, $r_0$ scales with the cluster temperature as 
$r_0 \backsimeq 0.50\,T^{1/2}$~Mpc where $T$ is in keV (see Fig.~16
in \cite{astro-ph/0507092}). For the best-fit temperature of $T=5.5$keV derived from the global spectrum,  $r_0 = 1.17$keV.  We fixed $r_0$ at the value suggested 
by the low redshift clusters only with an additional redshift scaling $r_0 \propto 1/E(z)$,
where $E(z) = \sqrt{\Omega_M (z+1)^3+\Omega_{\Lambda}}$,
to account for the redshift dependence of the virial radius for
clusters with a fixed temperature \citep{1998ApJ...495...80B}.
To fit the observed 
profiles, we projected the 3-D model along the line
of sight using the emission measure profile from the best fit
$\alpha$-$\beta$ model. Projection was based on a weighting method which
correctly predicts the best-fit spectral $T$ for a mixture of different
temperature components \citep{2004MNRAS.354...10M,astro-ph/0504098}.

Our 3-D model  has a flat shape in the central region ($r < 0.6 r_0 / \gamma ^ {0.5}$) and 
is not intended for modeling  central temperature drops associated with a cooling region.
The innermost bin shows an apparent temperature drop and can indicate the presence of
a cooling flow. Therefore  we excluded the central bin  from the fit. 
Taking into account the fact that the temperature profile was corrected for the XMM PSF 
this procedure is correct.  The obtained
best-fit model,  the deconvolved  and original temperature
profiles are shown in Fig.~\ref{oneplot}. The uncertainties on 
the best-fit model  were calculated from Monte-Carlo simulations. 
We applied Gaussian scatter to the observed temperature profile within its uncertainties 
and fitted simulated profiles. The best-fit model uncertainties estimated as 
\emph{rms} scatters in narrow radial bins from 1000 simulations are shown in Fig.~\ref{oneplot}..

\subsection{Mass and Luminosity measurements}

Assuming hydrostatic equilibrium for the ICM, we can use the best fit
temperature and density profiles to derive the total cluster masses:
\begin{equation}
  \label{eq:def1_m}
M(r)=-\frac{rT(r)}{G \mu m_p}\left( \frac{d \log \rho(r)}{d \log r} + \frac{d \log T(r)}{ d \log r} \right)
\end{equation}

We calculated the total mass of $M_{500}=(5.1 \pm 1.1) \times 10^{14} M_{\odot}$ and 
the radius of $r_{500}=1.09 \pm 0.08$Mpc  corresponding 
to the  mean overdensity $\Delta=500$ relative to the critical density
at the cluster redshift. The uncertainties on the total mass due to the temperature 
profile were calculated  
analogically to the uncertainties on the best-fit temperature profile model. 
We calculated the total mass for each simulated profile. Then the uncertainties were 
calculated as the boundaries of the region containing $90\%$ of all realizations.
The total mass uncertainties due to the error on
the density gradient  were calculated following \cite{Pratt02}.
The value of $d \log \rho(r) / d \log r$ at $r_{500}$  was considered  as  
an independent parameter of  the $\alpha$-$\beta$ model instead of $\beta$. 
We refit the surface brightness profile with the new parameter set and 
measured the uncertainties on $d \log \rho(r) / d \log r$ at $r_{500}$.
The final total mass uncertainties were calculated by adding quadratically  
the total mass error due to the temperature profile and the density profile.
The uncertainties on $R_{500}$ are related to the total mass uncertainties as
$\sigma_{M_{500}} / M_{500} = 3 \sigma_{r_{500}} / r_{500}$.

We calculated the emission measure-weighted
temperatures (volume-averaged with weight  $w = \rho _g ^ 2$) of $T_{emw}=6.0 \pm 0.7$  within 70kpc $<$ $r$ $<$ $r_{500}$. It is not a surprise that obtained $T_{emw}$ is higher than the best-fit 
spectroscopic temperature because we excluded  the innermost bin from the temperature profile fit.
The uncertainties for $T_{emw}$ were calculated from Monte-Carlo simulations in the similar way 
as for the temperature profile best-fit.

We can compare the obtained $M_{500}$ with  predicted value from $M-T$ relations. 
We will use as a low-redshift reference the $M-T$ relation from \cite{astro-ph/0507092},
which is similar to the $M-T$ relation measured by \cite{2005astro.ph..2210A}.
The $M-T$ relation derived by \cite{astro-ph/0507092} predicts 
$M_{500}= 6.1 \times 10^{14} M_{\odot}$ for a 6 keV cluster.
The self-similar theory predicts that for the same temperature the mass evolves 
as $M_{\Delta} \propto E(z)^{-1}$ with $z$, where $E(z)=H(z)/H_0=(0.3 (1+z)^3+0.7)^{1/2}$ 
(e.g. \cite{1998ApJ...495...80B}).  Dividing the predicted mass by $E(z=0.29)$ to place it
at $z=0.29$, we obtain $M_{500}= 5.3 \times 10^{14} M_{\odot}$. This value is in 
agreement with the derived  $M_{500}=(5.1 \pm 1.0) \times 10^{14} M_{\odot}$.

We derived the unabsorbed bolometric luminosity of $L_x = (8.4 \pm 0.5) \times 10^{44} \, h^{-2}_{71}
\, erg \, s^{-1}$ using the Mewe-Kaastra-Liedahl plasma emission model. We used $T_{emw}$ as 
the temperature parameter for the model. The model was normalized by the following count rate:
we subtracted from the observed $0.8-2.5$keV count rate calculated within $0 < r < 1.4$Mpc 
the count rate calculated from the best-fit $\alpha$-$\beta$ model within $r < 70$kpc, and multiplied
the result by 1.06 \citep{Markevitch98}. 

The fact that our analysis is similar to the one done by \cite{Markevitch98} for low redshift
 clusters allows us to compare 
the obtained luminosity with the prediction from  their $L-T$ relation. 
The $L-T$ relation derived by \cite{Markevitch98} predicts the luminosity of 
$6.2\times 10^{44} \, h^{-2}_{71} \, erg \, s^{-1}$ for a 6 keV cluster.
It is well below the derived luminosity. However, correcting 
the predicted luminosity for the evolution in $L-T$ relation, $L_{z}=(1+z)^{1.5} L_{0}$, reported
by \cite{Vikhlinin02}, we obtain  $9.2\times 10^{44} \, h^{-2}_{71} \, erg \, s^{-1}$.
This value is close to the observed one.

\section{Discussion and Conclusions}

Our deep XMM-Newton observations of M31 have shown that
RX J0046.4+4204 is not located in that galaxy, but rather is actually
a distant cluster of galaxies. We found RX J0046.4+4204 has spatially
extended X-ray emission and that the spectrum clearly shows a
red-shifted iron emission line. Straightforward fitting of the iron
line yields a cluster redshift of Z=0.290 with 2$\%$ accuracy and that
the redshift values estimated independently from EPIC-PN and MOS data
are in good agreement within measurement errors.

The large scale spatial distribution of RX J0046.4+4204 is well fit 
by the $\alpha$-$\beta$ model  with $\beta = 0.70 \pm 0.8$, a
core radius $r_c = (56 \pm 16 )\arcsec$ or  $r_c = (240  \pm  69)$kpc for $z$=0.290
, and $\alpha = 1.54 \pm 0.25$. The obtained values of
$\beta$ and $r_c$ are consistent with the parameters of typical clusters \citep{1999ApJ...525...47V}. 
The derived $\alpha$ agrees with values measured for cooling flow clusters \citep{astro-ph/0507092}.

The spatially integrated X-ray continuum is well fit by red-shifted
(z=0.290) Mewe-Kaastra-Liedahl plasma emission model with low energy photo-electric
absorption. The best fitting  global model for the joint PN, MOS1, and MOS2 measurements 
yields the parameters $kT(keV) = 5.5 \pm 0.5$, fractional solar abundance $Z_{\odot} = 
0.57 \pm ^{0.15} _{0.13}$,
,a redshift $z = 0.290 \pm 0.005 $, and a column depth
$n_{H} = 2.2 ( \pm ^{0.2} _{0.2} ) \times 10 ^ {21} cm ^ {-2}$. This derived
column depth is significantly larger than the Galactic hydrogen column 
in the direction of M31, $\sim 7   \times 10 ^ {20} cm ^ {-2}$ \citep{1990ARA&A..28..215D}.

It is interesting to compare the value of absorption obtained for the
cluster  with absorption for the nearby X-ray sources. Fig~\ref{image}({\em Right}) shows
five bright X-ray sources lying in the direct vicinity of RXJ0046.4+4204 for
which \cite{astro-ph/0401227} were able to measure column depth based on 
their X-ray spectra and to identify some of them.
XMMU J004540.5+420806(source $\#5$ in Fig. \ref{image}) was identified as a foreground
 star with the column depth of 
$0.2 \times 10 ^ {21} cm ^ {-2}$ (two sigma upper limit). 
XMMU J004648.0+420851(source $\#3$) was identified as a background radio 
source with the column depth of 
$(4.0 \pm ^{1.0}_{5.0}) \times 10 ^ {21} cm ^ {-2}$. XMMU J004627.0+420151(source $\#1$) 
was identified as a globular cluster source in M31. The derived column 
depth  for this source was $(1.3 \pm 0.1) \times 10 ^ {21} cm ^ {-2}$. The nature of two 
last sources, XMMU J004611.5+420826(source $\#4$) and XMMU J004703.6+420449
(source $\#2$), was unclear. 
\cite{astro-ph/0401227} proposed that these sources could be two AGN located in the background 
of M31. The derived column depths of XMMU J004611.5+420826 and XMMU J004703.6+420449 were 
$(2.5 \pm 0.3)\times 10 ^ {21} cm ^ {-2}$ and $(2.3 \pm 0.6) \times 10 ^ {21} cm ^ {-2}$, 
respectively. 

The absorption value of the background radio source is 2 times higher than 
the cluster absorption and most likely to be intrinsic. 
The column depth of the globular cluster candidate is smaller than that 
of the cluster. To explain this fact, it might be proposed that the globular 
cluster candidate 
is located in front of the disk of M31, while the cluster is obscured by the disk. 
On the other hand, XMMU J004611.5+420826 and XMMU J004703.6+420449 have column depths consistent 
with that of the cluster, suggesting that they could be also located in the background of M31, 
although it is unclear what fraction of the column depths is intrinsic.
This interpretation is in general agreement with results of spectral fitting of a large sample 
of  M31 globular cluster X-ray sources \citep{M31_GCS}.
\cite{M31_GCS} found that globular cluster sources located in front
of M31 disk have typical values of absorbing column in the range of 
$(0.5 - 1.5)\times 10^{21}$, while the sources located behind 
the disk or embedded into it show higher absorbing columns of $(2 - 4)\times 10^{21}$ 
cm $^{-2}$.

The extracted temperature profile corrected for the XMM PSF  shows the central 
temperature decline 
that confirms the indication of cooling flow presence  from the spatial analysis. Using 
the spatially resolve temperature profile we derived $R_{500}=1.09 \pm 0.07$Mpc, 
$M_{500}= (5.3 \pm 1.0)\times 10^{14} M_{\odot}$, and $T_{emw}=6.0 \pm 0.7$  within $R_{500}$. The values 
of $M_{500}$ corrected for the evolution and $T_{emw}$ are in agreement with local $M-T$ relations.

The study we have presented here shows the utility of sensitive X-ray
observations for identifying and studying clusters of galaxies in directions where
foreground confusion or heavy optical extinction makes optical selection complicated. 

\acknowledgments 
We thank A. Vikhlinin for extensive discussions and helpful comments. 
Also we'd like to thank the referee for providing useful suggestions.
This paper is based in part on observations obtained with {\em
XMM-Newton}, an ESA science mission with instruments and contributions 
directly funded by ESA Member States and the USA (NASA). This work was 
supported in part by NASA grant NAG5-12390 and by Internal Laboratory 
Directed Research and Development funding at Los Alamos National Laboratory.

\end{document}